\documentclass[preprint2]{aastex}

\slugcomment{Accepted to The Astrophysical Journal}

\shorttitle{Color Identification of GRB Afterglows}
\shortauthors{Rhoads}

\begin{document}
\def\mag{\,\hbox{mag}}
\def\micron{\mu m}
\def\uvvi{$U-V$,~$V-I$}
\def\bvri{$B-V$,~$R-I$}
\def\pc{pc}
\def\ee{{\cal E}}
\def\day{\,\hbox{day}}

\title{Snapshot Identification of Gamma Ray Burst Optical
Afterglows\footnote{Dedicated to the memory of Jan van Paradijs.}}

\author{James E. Rhoads}
\affil{Space Telescope Science Institute}

\begin{abstract}
Gamma ray burst afterglows can be identified in single epoch
observations using three or more optical filters.  This method relies
on color measurements to distinguish the power law spectrum of an afterglow
from the curved spectra of stars.  Observations in a fourth filter
will further distinguish between afterglows and most galaxies up to
redshifts $z \approx 1$.  Many afterglows can also be identified with
fewer filters using ultraviolet excess, infrared excess, or Lyman
break techniques.
By allowing faster identification of 
gamma ray burst afterglows, these color methods will increase the fraction
of bursts for which optical spectroscopy and other narrow-field
observations can be obtained.    
Because quasar colors can match those of afterglows, the maximum error box
size where an unambiguous identification can be expected is
set by the flux limit of the afterglow search and the
quasar number-flux relation.  For currently typical error boxes (10 --
100 square arcminutes), little contamination is expected at magnitudes
$R \la 21.5 \pm 0.5$.  Archival data demonstrates
that the afterglow of GRB 000301C could have been identified using this
method.  In addition to finding gamma ray burst counterparts, this
method will have applications in ``orphan afterglow'' searches used to
constrain gamma ray burst collimation.
\end{abstract}


\section{Introduction}
The search for counterparts to gamma ray bursts (GRBs) at longer
wavelengths has been recognized from the beginning as a key to
understanding the bursts' nature.  This motivated models that
predicted gamma ray burst afterglows from hot material in
relativistically expanding GRB remnants (Paczy\'{n}ski \& Rhoads 1993;
Katz 1994; M\'{e}sz\'{a}ros \& Rees 1997).  The afterglow was
predicted to arise from synchrotron radiation, and its spectrum is
therefore well described by a broken power law.  Both the prediction
of fading counterparts with power law spectra and the general
expectation that GRB counterparts would tell us a great deal about the
bursts have been amply borne out by observation since the first
successful observations of GRB afterglows in 1997 (Costa et al 1997;
van Paradijs et al 1997; Frail et al 1997).

The greatest rewards so far have come from detailed studies of
individual afterglows with the best time and spectral coverage.  Rapid
afterglow identifications are a key to achieving such data sets,
because some essential types of data cannot be obtained before the
afterglow position is known to $\sim 1$ arcminute (near-infrared and
submillimeter photometry) or even $\sim 1$ arcsecond (spectroscopy).
Traditionally, afterglows are found through a variability search
within GRB error boxes.  This requires multiple epoch observations
whose separation in time $\Delta t$ must be a reasonable fraction
($\ga 0.3$) of the time elapsed from the GRB to the first epoch.  This
lag is usually further increased by telescope scheduling issues to
$\sim 1$ day.  Moreover, afterglow light curves occasionally include
``plateau'' phases (e.g. GRB 970508 at $t < 2$ days [Pedersen et al
1998]; GRB 971214 at $t < 0.3$ days [Gorosabel et al 1998];
and GRB 000301C at $t\sim 6$ days [Bernabei et al 2000; Masetti et al
2000]).
Two-epoch observations of such afterglows could miss them entirely.

This delay in afterglow identification is not necessary, because
afterglows are characterized not only by fading behavior but also by
power law spectra.  Such spectra are relatively rare in the optical
sky, which is dominated by Galactic stars at bright flux levels and by
galaxies at faint flux levels.  Relative to a stellar spectrum, a GRB
afterglow is blue at blue wavelengths ($\la 0.4 \micron$), and red at
red wavelengths ($\ga 0.6 \micron$).  A color-color plot should
therefore isolate a GRB afterglow from nearby stars efficiently,
provided it spans a sufficient wavelength range to detect spectral
curvature (or the Balmer jump) in stars of any surface temperature.
Under many circumstances it will also be possible to identify
afterglows using simpler though less robust single color criteria
(either ultraviolet excess [``UVX''] or infrared excess [``IRX'']).
Moreover, absolute photometric calibration is not required so long as
a sufficient solid angle is observed to empirically determine the
stellar locus in color space.

Section~\ref{col_mod_results} describes model color-color plots that
show the expected loci of GRB afterglows, stars, galaxies, and
quasars.  Section~\ref{empirical} demonstrates empirically that the
color-color method could have identified the afterglow of GRB 000301C
in a single epoch.  Finally, section~\ref{discussion} considers
advantages and limitations of the method, including confusion limits
as a function of flux level, likely selection effects in the sample of
afterglows found through color-based searches, and applicability of
the method to ``orphan afterglow'' searches.

\section{Color-Color Space Models} \label{col_mod_results}

To explore this method in detail, I have calculated synthetic color
space locations for GRB afterglows, stars, galaxies, and quasars.
Model calculations were performed for the widely available Johnson U,
B, and V filters and Cousins R and I filters, though color-color
searches will work for any set of filters spanning a sufficient
wavelength range.  Filter transmission curves were taken from Bessell (1990).
The photometric zero point for all Johnson-Cousins system synthetic
colors was set to the Vega model atmosphere of Kurucz (1979).

Model spectra of objects at high redshift were attenuated using the
mean intergalactic absorption given by Madau (1995)\footnote{The
photoelectric absorption was treated using the approximation to
Madau's equation~16 from his footnote~3, which is within 5\% of the
full calculation.}. 
For redshifts $z \ga 2$, the flux in the bluest filter can
be greatly reduced by neutral hydrogen absorption in the intergalactic
medium.  This absorption consists of a superposition of Lyman $\alpha$
forest lines, higher order Lyman lines, and bound-free Lyman continuum
absorption (Madau 1995).  The net effect is a
shift in the bluest color measured.  
The relevant redshift depends on the choice of bluest filter.  The Lyman
$\alpha$ line is redshifted to the peak transmission wavelength of the
Johnson U band at $z=2.04$, and the Johnson B filter
around $z=2.5$.


In addition, I calculated reddening vectors for dust both in the Milky
Way and in afterglow host galaxies at a range of redshifts, using the
empirical fits of Pei (1992) to the extinction laws of the Milky Way
and the Magellanic Clouds.

In the following subsections, I describe the color calculations for
each class of object considered, and then summarize the results.

\paragraph{Stars:}
Stellar colors were derived from the spectrophotometric catalog of
Gunn and Stryker (1983), which spans spectral types from O to M and
luminosity classes from the main sequence to supergiants.

\paragraph{GRB Afterglows:}
Afterglows were taken as pure power law spectra, $f_\nu \propto
\nu^{\alpha}$, with indices $0.5 \la -\alpha \la 1.5$.  Such spectra
describe synchrotron emission from a power law distribution of
electrons away from break frequencies.  Deviations from power law
spectra occur near spectral breaks (see Sari, Piran, \& Narayan 1998)
and in the case of interstellar absorption (section~\ref{discussion}).
The range of
spectral indices considered corresponds to the commonly measured range
of electron energy indices, $2 < p < 3$. Here $p$ describes the
number-energy distribution of electrons accelerated at the GRB
remnant's external shock, $N(\ee) \propto \ee^{-p}$, and I have
considered spectra both above and below the cooling frequency.

\paragraph{Galaxies:}
Galaxy spectra were taken from the GISSEL stellar population synthesis
code (Bruzual \& Charlot 1993) for a few representative star formation
histories. 
No attempt was made to include galaxy evolution effects, which may result
in elliptical galaxy colors that are slightly too red at high redshift
(see, e.g., Bruzual [1996]).
The resulting galaxy colors agree with those
computed by Fukugita, Shimasaku, \& Ichikawa (1995) to
within $0.1 \mag$ for most filters, galaxy spectral types, and
redshifts; this is acceptable given the differences in the input
spectra in the two works.

Galaxy spectra at optical wavelengths are a superposition of diverse
stellar spectra, modified slightly by nebular emission and dust effects.
This superposition implies that galaxies have less
sharply peaked spectra than individual stars and so lie on the same side of the
color-color space stellar locus as power law spectra.  Indeed,
the spectrum of a star forming galaxy can be a good
power law over most of the optical window.
Fortunately, the $4000$\AA\ break due to stellar atmospheric features
interrupts this approximate power law and can
be used to distinguish almost all galaxies at $z \la 1$ from afterglow
candidates given sufficient color information.  
With observations in three filters,
there is some redshift where the $4000$\AA\ break lies in the middle
filter, and galaxy colors can match those of a pure power law.  The
addition of a fourth filter breaks this degeneracy.  For example, in
the $U-B$, $R-I$ plane, an irregular galaxy at $z \approx 0.4$ lies on
the power law sequence; however, its $B-R$ color is much redder than
the corresponding power law.
In addition to color criteria, it is of course possible
to distinguish galaxies from GRB afterglows morphologically.
The fraction of galaxies that appear extended depends on the
available spatial resolution, with the a few galaxies appearing
pointlike even in Hubble Space Telescope images, but subarcsecond
resolution will in practice greatly reduce confusion due to galaxies.

\paragraph{Quasars:}
Quasar colors were calculated using the composite quasar spectrum by
Francis et al (1991), extrapolated to $\lambda > 0.6\micron$ with
$f_\lambda \propto \lambda^{-2.18}$.  This extrapolation will result
in modest color errors for low redshifts where the H$\alpha$ line
falls in a measured filter, but should not affect our conclusions
greatly.

Quasar spectra are reasonably well approximated by power laws (plus
broad emission lines) across a wide wavelength range.
Color-based criteria have been used to identify quasars for nearly 40
years (see the review by Burbidge 1967), and the use of far-red
filters to supplement bluer UBV passbands was suggested as early as
1967 (Braccesi 1967; Braccesi, Lynds, \& Sandage 1968).  Quasars are
therefore extremely difficult to distinguish from GRB afterglows using
only broad band colors.  The main discriminant available is that
quasars tend to be somewhat bluer than typical afterglows.
In practice, we expect confusion with quasars to set the
maximum area over which an unambiguous afterglow identification can be
expected.  This confusion can be resolved with low resolution spectra,
which identify most quasars unambiguously by their strong, broad
emission lines.

\subsection{Results}
\begin{figure}
\epsscale{0.8}
\plotone{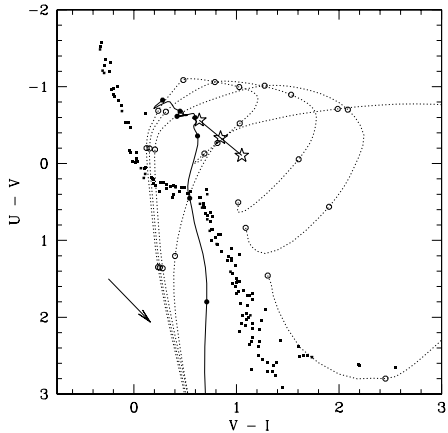}
\caption{Model U-V, V-I color-color plane.  5-pointed stars mark the
location of GRB afterglows, represented by pure power law spectra with
indices $f_\nu \propto \nu^{\alpha}$ for $\alpha = -0.5, -1.0, -1.5$.
Small points represent stars from the Gunn-Stryker (1983) atlas.
Dotted lines show the locations of different galaxy spectra as a
function of redshift.  Star formation activity increases from lower
right (Elliptical galaxy model) through spiral models (Sb and Sc) to
an actively star forming Magellanic irregular (Im) model.  Open
circles along these tracks mark intervals of 0.5 in redshift, from
zero to 3.0.  No galaxy evolution is incorporated in these tracks.
Finally, the solid line shows the locus of quasars as a function of
redshift.  Filled points along this track again mark redshift
intervals of 0.5 from zero to 3.0.  Intergalactic absorption will
shift the afterglow locus to redder $U-V$ at fixed $V-I$ for $z \ga
2.0$.
\label{mod_uvvi}}
\end{figure}

\begin{figure}
\epsscale{0.8}
\plotone{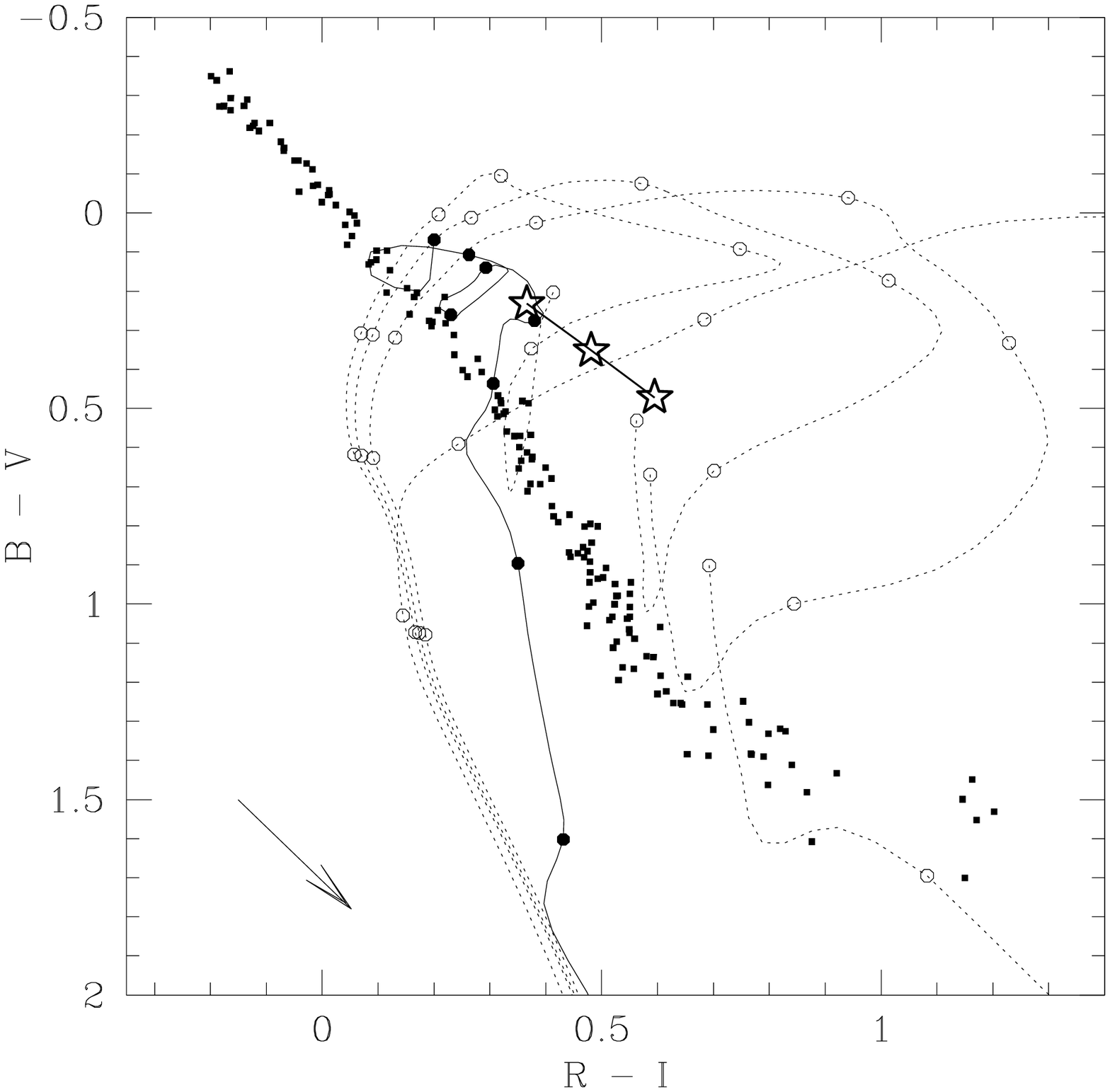}
\caption{Model B-V, R-I color-color plane. 
Symbols are the same as for figure~\ref{mod_uvvi}.  
Because neutral hydrogen absorption first becomes important at higher
redshift in the B band than the U band, color tracks for
galaxies and quasars extend to somewhat higher redshift here, and the
afterglow locus shifts to redder $B-V$ for $z \ga 2.5$.
\label{mod_bvri}}
\end{figure}

Sample color-color diagrams are shown for the \uvvi\ plane in
figure~\ref{mod_uvvi}, and the \bvri\ plane in figure~\ref{mod_bvri}.
These have been chosen as representative cases where the two-color
method works.
Additional combinations of filters are also effective provided that at
least three filters are used, the bluest filter is at least as blue as
the Johnson B band, and the reddest is at least as red as Cousins I.
Unfortunately, the observationally easiest filters (BVR) do not yield
an adequate separation between afterglow and stellar colors to expect
reliable identification of afterglows, barring unusually high
photometric precision.

Given a sufficient filter set, the afterglows are
well separated from the stellar locus. 
%
Moreover, the reddening vector for Milky Way dust runs essentially
parallel to both the stellar sequence and the power law spectral
sequence.  Thus, even though reddened afterglow spectra are no longer
strict power laws, they remain distinct in color-color space.

Of greater concern is dust in the GRB host galaxy.  Because this dust is at
high redshift, its $2175$\AA\ feature can enter the observed wavelength
range, changing the direction of the reddening vector
in color-color space.  For redshifts $2.2 \la z \la 3.3$, this feature
falls in the reddest filter of the color-color plot, thus
making the long-wavelength color (e.g., V-I) slightly {\it bluer}.  
A moderate amount of dust ($A_V = 0.5$) can then
move the afterglow locus onto the stellar locus in color-color space.
However, this effect is only important for Milky-Way type dust.
The most important consequence of dust for
color-based afterglow searches is therefore not reddening but
extinction, which can reduce the received flux below the detection
limit of the search whatever the reddening law.  However, GRBs can
destroy dust at distances up to $10~\pc$ (Waxman \& Draine 2000) or
beyond (Fruchter, Krolik, \& Rhoads 2001), reducing these concerns
substantially for bursts at high Galactic latitude.

\section{Empirical Tests} \label{empirical}
Archival multicolor data is available for a few GRB fields from the
U.S. Naval Observatory's GRB followup team (see Henden et al 2000).  I
have used this data for the field of GRB~000301C, together with
published photometry of the afterglow (Jensen et al 2000), for an
empirical demonstration of color-based optical afterglow searches.

\begin{figure}
\epsscale{1.01}
\plottwo{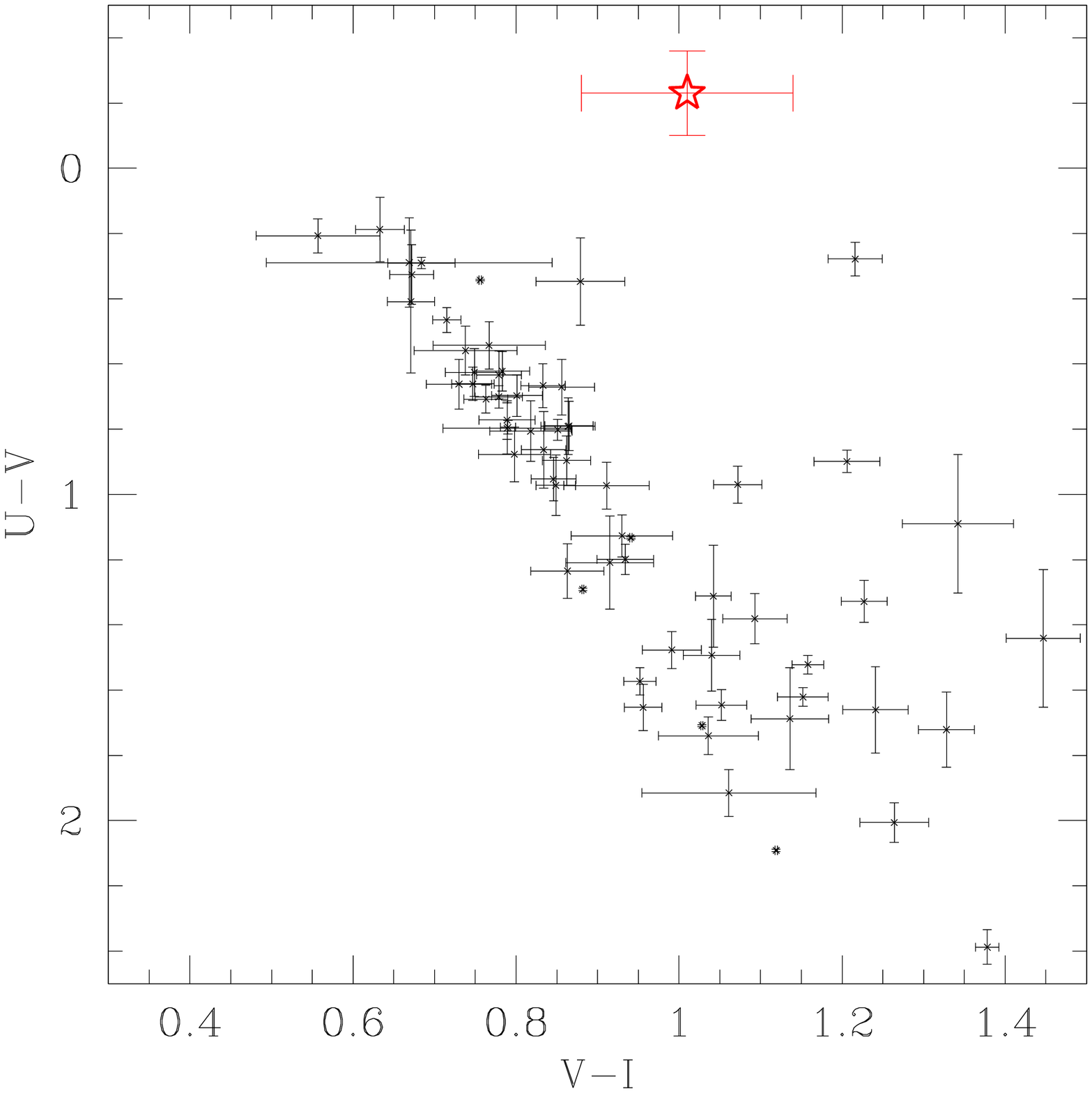}{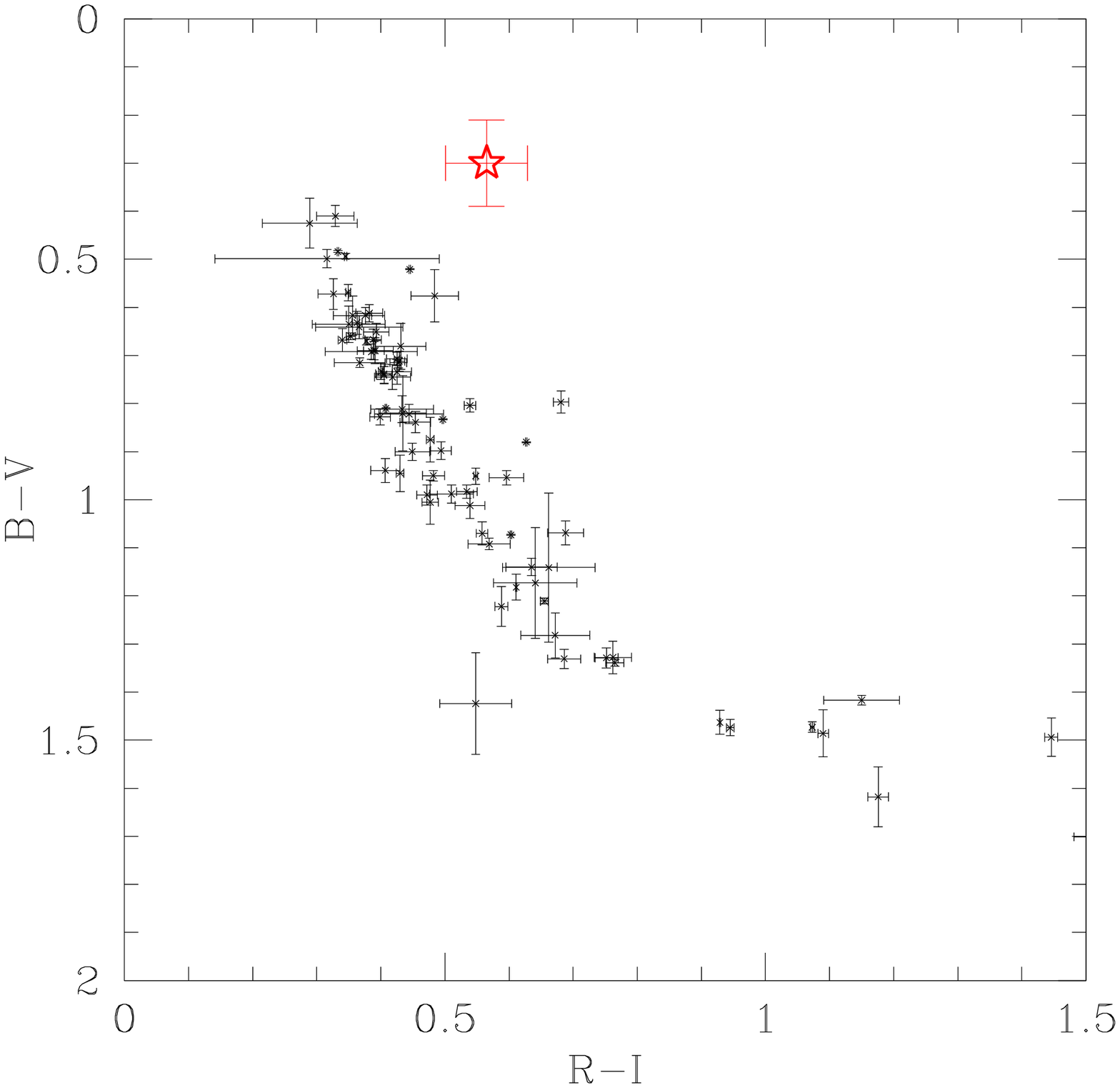}
\caption{Color-color plots for the field of GRB 000301C.  Data are
taken from Jensen et al (2000) for the GRB afterglow (marked by a
five-pointed star symbol); and from Henden et al (2000) for other
objects in the field.  For the Henden et al data, error bars on U-V
were computed by adding in quadrature the reported U-B and B-V errors;
similarly, V-I errors were computed from V-R and R-I.  The GRB is an
outlier in both plots, and the most dramatic outlier in the B-V, R-I plane.
\label{301c}}
\end{figure}

Two color-color plots for this field are shown in
figure~\ref{301c}.  In both cases, we
see that the afterglow is an outlier in color-color space.  In the
\uvvi\ figure, there are a few other outliers, all of which are
systematically redder than the afterglow in both colors.  In the \bvri\ 
plot, the afterglow is the single most dramatic outlier and would be
the first target chosen for prompt spectroscopy or other rapid
narrow-field followup observations.

\section{Discussion}\label{discussion}
Color-based searches have the potential to identify GRB afterglows
faster than a two-epoch variability search, and hence to allow
more thorough followup of many afterglows.
There are three basic color selection criteria available for
afterglows: IR excess (``IRX''), UV excess (``UVX''), and Lyman break.
The appropriate combination of these depends on both the redshift of
the burst and the available instrumentation.  The primary focus of
this paper is on color-color plane methods, which combine the IR
excess and UV excess criteria and are therefore more robust than
either criterion alone.

Still, as the observations of GRB 000301C demonstrate, potentially
occupied regions of color-color space remain vacant in practice for
high-latitude fields observed to a limiting magnitude $V\sim 21$.  In
particular, there are no very hot stars, few or no high redshift
galaxies, and few or no quasars.  Under these circumstances, a simple
UV Excess (``UVX'') test could identify the afterglow on the basis of
a single color measurement (preferably including one filter at
$\lambda < 4000$\AA).

Likewise, at near-infrared wavelengths, most stars and galaxies will
have spectral slopes near the Rayleigh-Jeans value ($2$) and will
therefore be bluer than afterglows.  This is again demonstrated by GRB
000301C, for which we found $J-K' \approx 1.45$ (Rhoads \& Fruchter
2001), about $0.3$ magnitudes redder than even the coolest stars tabulated
by Johnson (1966).  However, UVX and IRX methods are vulnerable to the
presence of very hot stars and highly reddened sources respectively,
while the two color method is not.

The Lyman break (or ``dropout'') method uses intergalactic absorption
to find high redshift afterglows whose strongly attenuated blue flux
places them in a region of color space occupied by essentially all
classes of high redshift object, on the opposite side of the stellar
locus from low redshift afterglows.  The observational requirements to
find these bursts from Lyman break colors are more exacting.  Long
integrations in the bluest filter are required to conclusively
identify ``dropout'' afterglows, while lower redshift afterglows are
unusually bright in the bluest filter and require correspondingly
shorter integrations.  Even so, most afterglows are brighter than
their host galaxies at discovery, which means that a high redshift
afterglow may be an easier dropout to identify than a high redshift
galaxy.

Color-based searches are subject to certain selection effects, which
must be considered in interpreting their results.  These effects
include:
(1) Redshift: The onset of intergalactic hydrogen absorption in the
bluest filter observed determines the highest redshift where UV excess
methods can work and the lowest redshift where Lyman break methods can
work.  The UVX method will fail at $z\approx 2$ for U band and
$z\approx 2.5$ for B band, while the Lyman break method first becomes
practical in the U band at $z\approx 3.3$.  Thus, two-color optical
searches will not identify afterglows between about $z=2.5$ and
$z=3.3$.  The IR excess will continue to be practical to quite high
redshift, e.g. to $z\approx 15$ for a $J-K$ color, since the onset of
neutral hydrogen absorption simply accentuates an IR excess until the
absorption reaches the redder filter.
Because there is considerable variance in the hydrogen column density
along different lines of sight (Madau 1995), the redshift
boundaries discussed here are not sharp demarcations.  Rather, 
the detection efficiency declines smoothly, and
the characteristic redshifts given above correspond to about $50\%$ of
peak efficiency.

(2)
Heavily obscured afterglows can drop below the search's detection threshold.
Extinction affects any search method, though it is exacerbated for UVX
and two-color methods by the increase of extinction towards bluer
optical-UV wavelengths and the necessity of observations in observer
frame blue light.  Additionally,
there will be a mild selection against finding afterglows in
galaxies with Milky Way type dust extinction at redshifts $2.2 \la z
\la 3.3$ using two-color optical searches.  However, this overlaps the
stronger selection due to intergalactic Hydrogen absorption to such an
extent as to be almost unimportant.

(3) For gamma ray bursts occurring near dense molecular gas,
absorption by excited molecular hydrogen can occur for rest
wavelengths as long as $1650$\AA\ (Draine 2000).  Absorption between
$1650$ and $1300$\AA\ can reach 50\% for H$_2$ column densities around
$10^{19}$ cm$^{-2}$.  This will act to exclude afterglows in molecular
clouds from UVX or two-color samples at redshifts beyond $1.3$ (using U band)
or $1.65$ (using B).

(4) Because afterglow spectra are broken power laws rather than pure
power laws, there will be times in the evolution of each afterglow
when its colors deviate from the models in
section~\ref{col_mod_results}.  Breaks in synchrotron spectra
are not absolutely sharp, so this is unlikely to be a major selection
effect for purely optical searches, where the wavelength range spanned
is only a factor of $\sim 2.5$.  Searches employing larger wavelength
coverage (UV or IR) will be more sensitive to this effect.

(5) Afterglows located atop brighter host galaxies may be
hard to find because the dominant color is that of the galaxy.  This
effect will depend on instrumental spatial resolution as well as the
intrinsic properties of the afterglow and host galaxy.  (Variability
searches can be similarly compromised unless image subtraction methods
are used.)

(6) An advantage of this method is that short-term ``plateaus'' in afterglow
light curves do not affect it.  It may therefore find some afterglows
that traditional variability searches miss.

Even afterglows that lie in the predicted region of color space 
may be difficult to identify by these methods in cases where the
error box is so large that a substantial number of other sources
occupy the same region of color space.  The critical error box size
where one ``confusing'' source is expected depends on the magnitude
limit of the search.  Ultimately, the quasar number-flux relation
gives a lower limit to the confusion level.  Additionally, for three
filter searches, star forming galaxies at $z \sim 0.5$ can add to the
confusion.
%
%
%
To achieve a unique afterglow identification, it is necessary to match
the magnitude limit of the search to the size of the error box, and to
observe the field while the afterglow remains brighter than this
magnitude limit.  A unique identification may not always be needed,
however.  In particular, a list of several viable candidates may be
enough to obtain a same-night afterglow spectrum using a multiobject
spectrograph with a sufficient field of view and rapid setup
procedure.  The correct candidate could then be identified later based
on either spectra or variability.

The quasar number-magnitude relation for $z<2.2$ and magnitude $B<24$
is predicted by Malhotra \& Turner (1995) for both ($\Omega=1$,
$\Lambda=0$) and ($\Omega=0.1$, $\Lambda=0.9$) cosmologies, based on
observations to $B<22$ by Boyle et al (1987) and Boyle, Shanks, \&
Peterson (1988).

\begin{figure}
\epsscale{0.8}
\plotone{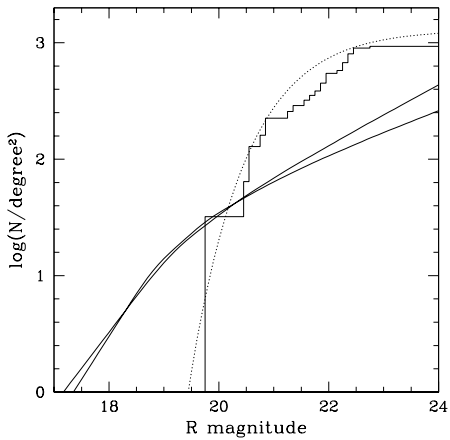}
\caption{Cumulative number-magnitude relation for sources whose optical colors
resemble GRB afterglows.  The solid lines shows quasar counts
predicted by models A and B of Malhotra \& Turner (1995).  The dotted
line shows the predicted number counts for galaxies with $0.4 < z <
0.6$ using the locally derived luminosity function from Loveday et al
(1992).  Counts of blue galaxies should be $\sim 30\%$ lower.  The
histogram shows the Canada-France Redshift Survey
data for galaxies with $0.4 < z < 0.6$ and
with $V-I$ colors in the afterglow range (Lilly et al 1995a; Le Fevre
et al 1995; Hammer et al 1995).  The total number of galaxies around
22nd magnitude is a factor $\sim 10$ above the galaxy counts shown.
The plotted curves can be taken as an approximate guide to the number
of potentially confusing sources expected for a given area searched
and limiting magnitude attained.  The exact number can depend
substantially on additional factors, in particular photometric
accuracy, spatial resolution, and the availability of a third
independent color to distinguish $z \sim 0.5$ galaxies from
afterglows.
\label{foregrounds}}
\end{figure}

%
The approximate range of redshift where star forming galaxies overlap
the GRB afterglow locus in color-color space is $0.35 \la z \la 0.6$
for the $U-V$, $V-I$ diagram, or $0.45 \la z \la 0.6$ for the $B-V$,
$R-I$ diagram.  I estimate a number-magnitude relation for such
galaxies using both the local luminosity function from Loveday et al
(1992) and direct counts of galaxies with $0.55 < V-I < 1.28$ and $0.4
< z < 0.6$ from the Canada-France Redshift Survey (Lilly et al 1995a;
Le Fevre et al 1995; Hammer et al 1995).  The predicted number counts
amount to approximately 10\% of all galaxy number counts (see, e.g.,
Lilly et al 1995b, Williams et al 1996) around $R=22$.  The quasar and
galaxy count estimates (converted to R band magnitudes using the
colors of an $f_\nu \propto \nu^{-1}$ power law) are plotted together
in figure~\ref{foregrounds}.  We see that quasars are the more
important foreground for $R<20$.  At fainter magnitudes, galaxies
appear more important, but this may be misleading since they can often
be identified using a third independent color and/or morphological
information.

Presently, most GRB error boxes have characteristic sizes of a few tens of
square arcminutes.  Thus, the brightest quasar expected in the error
box has a magnitude $R \sim 22$, while the brightest galaxy having
colors similar to an afterglow might be a magnitude brighter.  Both
this depth and the error box area are reasonably well matched to the
capabilities of moderate sized (1 to 2.5 meter) ground-based
telescopes.

\subsection{Optimizing the Observations}
To realize the promise of same-night afterglow identifications, a
rapid and standard data analysis procedure is desirable, and could be
aided by an observing procedure optimized for quick analysis.  
Archival bias frames and flatfield frames for all filters could be
useful, and precisely cospatial exposures in multiple filters (i.e.,
without offsets and with telescope guiding) could ease measurement of
object colors.

An additional constraint on observing strategy arises because the
target is likely to fade during the period of observations.  If the
multiple wavelengths cannot be observed simultaneously, it is then
better to begin with the bluest filter, followed by the reddest, and
only then move on to the middle filters.  This way, fading behavior
will accentuate the expected color offset rather than reducing it.
(The choice of bluest filter before reddest is driven by the
possibility that a spectral break may move through the observed
wavelength range during the course of observations.)

The optimal choice of filter set will depend on local conditions.  The
relative difficulty of U band observations is approximately balanced
by the large color-plane offsets obtained with U data ($\sim 0.7~\mag$
from the stellar locus), which relax the signal to noise
requirements considerably.

\subsection{Orphan Afterglow Searches}
If gamma ray bursts are highly collimated, we expect to observe many
more afterglows than GRBs (Rhoads 1997).  The search for such
``orphan afterglows'' is an important and largely model-independent
way of testing GRB collimation.

Color-based techniques will be a valuable addition to orphan
afterglow searches.  Afterglows remain rare phenomena even in
extreme collimation scenarios, and in order to find one it will be
necessary to survey a large area of sky that will contain many
quasars.  Thus, orphan afterglow searches must combine variability
with color tests to be reliable.  When enough filters are observed
simultaneously in a field where an archival first epoch is available,
it will be possible to identify new sources, test their colors for
consistency with a power law spectrum, and start intensive followup
of any viable afterglow candidates.  This will allow
much greater confidence in the identification of orphan afterglows
than is presently possible, since good spectra and light curves can
then be used to distinguish real afterglows from variable stars, high
redshift supernovae, active galactic nuclei, microlensing events, and
other optically variable sources.

As an illustrative example, consider Sloan Digital Sky Survey (SDSS;
see Knapp et al 1999).
The SDSS southern strip will cover an area $\sim 200$ degree$^2$ to
about 22nd magnitude in 5 broadband filters many ($\sim 45$) times.
The known GRB rate is $\sim 3$ per sky per day, and bright afterglows remain
above 22nd magnitude for a period $\sim 4$ days.  Thus the southern
strip has an effective coverage of $4 \times 10^4 \hbox{degree}^2
\day$.  The northern SDSS survey will cover $10^4$ degree$^2$ in a
single imaging epoch.  Fortunately, GRB afterglows would likely be
targeted for spectra based on their quasar-like colors, yielding a
second epoch.  The northern survey thus provides another $\sim 4
\times 10^4 \hbox{degree}^2 \day$.  Both parts together would be
expected to detect $\sim 6$ GRB afterglows.  If GRBs are tightly
collimated (opening angles $\la 10^\circ$), the expected number of
orphan afterglows rises to $\ga 25$.  By using the color space
properties to identify these afterglows, such a survey can place
strong limits on the collimation angle and hence energy requirements
of gamma ray bursts.

\section{Conclusions}
Color-based searches will allow faster identification of many gamma ray
burst afterglows.  Intensive followup of these afterglows can then
begin earlier and at brighter flux levels, yielding a larger sample of
bursts for which spectroscopic redshifts and well measured spectral
energy distributions (including roughly synoptic data at X-ray,
optical, near-infrared, submillimeter, and radio wavelengths) can be
obtained.  Such data sets are crucial to detailed studies of afterglow
physics.  Additionally, single-epoch color-based searches may yield
a more easily characterized statistical sample of afterglow detections
and nondetections than is currently available.  Such samples will
allow reliable studies of the population properties of gamma ray burst
afterglows, and thereby open the way for new insights into the nature
of gamma ray bursts.

\acknowledgements
I thank Sangeeta Malhotra for extensive discussions and comments on an
early draft of this paper; Andy Fruchter for suggesting a harder look
at IR excess methods; and Mauro Giavalisco, Arjun Dey, Norm
Grogin, and Richard Green for additional discussions.  Finally, I
thank Arne Henden and the USNO GRB team for providing freely the
photometric calibration data for GRB fields that allowed the first
empirical test of this method.  This work was supported by an
Institute Fellowship at The Space Telescope Science Institute
(STScI), which is operated by AURA under NASA contract NAS 5-26555.


\begin{references}
\reference{} Bernabei, S., Bartolini, C., Di Fabrizio, L., Guarnieri,
  A., Piccioni, A., \& Masetti, N. 2000, GCNC 599
\reference{} Bessell, M. S. 1990, \pasp\ 102, 1181 
\reference{} Boyle, B. J., Fong, R., Shanks, T., \& Peterson,
  B. A. 1988, \mnras\ 227, 717  
\reference{} Boyle, B. J., Shanks, T., \& Peterson, B. A. 1988,
  \mnras\ 235, 935  
\reference{} Braccesi, A. 1967, {\it Nuovo Cimento\/}, Ser. 10, 49, 148
\reference{} Braccesi, A., Lynds, R., \& Sandage, A. 1968, \apjl\ 152, L105
\reference{} Bruzual A., G., \& Charlot, S. 1993, \apj\ 405, 538
\reference{} Bruzual A., G. 1996, in ASP Conference Series 98, From
  Stars to Galaxies: The Impact of Stellar Physics on Galaxy Evolution,
  eds. C. Leitherer, U. Fritze-v.~Alvensleben, \& J. Huchra (San
  Francisco: ASP), 14
\reference{} Burbidge, E. M. 1967, \araa\ 5, 399 
\reference{} Costa, E., et al 1997, IAU Circular 6572
\reference{} Draine, B. T. 2000, \apj\ 532, 273 
\reference{} Frail, D. A., Kulkarni, S. R., Nicastro, L., Feroci, M.,
  \& Taylor, G. B. 1997, Nature 389, 261
\reference{} Francis, P. J., Hewett, P. C., Foltz, C. B., Chaffee,
  F. H., Weymann, R. J., \& Morris, S. L.  1991, \apj\ 373, 465.
\reference{} Fruchter, A. S., Krolik, J. H., \& Rhoads, J. E. 2001,
  submitted to \apj
\reference{} Fukugita, M., Shimasaku, K., \& Ichikawa, T. 1995, \pasp\
  107, 945    
\reference{} Gorosabel, J., et al 1998, \aap 335, L5
\reference{} Gunn, J. E., \& Stryker, L. L. 1983, \apjs\ 52, 121
\reference{} Hammer, F., Crampton, D., Le Fevre, O., \& Lilly,
  S. J. 1995, \apj\ 455, 88 
\reference{} Henden, A. A., and the USNO GRB team, GCN Circular 583, 2000
\reference{} Jensen, B. L., et al 2000, submitted to \aap; astro-ph/0005609
\reference{} Johnson, H. L. 1966, \araa\ 4, 193
\reference{} Katz, J. I. 1994, \apj\ 422, 248
\reference{} Knapp, G. R., Gunn, J. E., Margon, B., Lupton, R. H.,
  York, D., \& Strauss, M. 1999, {\it The Sloan Digital Sky Survey
  Project Book}, Version V1.3
\reference{} Kurucz, R. L. 1979, \apjsupp\ 40, 1
\reference{} Le Fevre, O., Crampton, D., Lilly, S. J., Hammer, F., \&
  Tresse, L. 1995, \apj\ 455, 60  
\reference{} Lilly, S. J., Hammer, F., Le Fevre, O., \& Crampton,
  D. 1995a, \apj\ 455, 75 
\reference{} Lilly, S. J., Hammer, F., Le Fevre, O., \& Crampton,
  D. 1995b, \apj\ 455, 108 
\reference{} Loveday, J., Peterson, B. A., Efstathiou, G., \& Maddox,
  S. J. 1992, \apj\ 390, 338 
\reference{} Madau, P. 1995,  \apj\ 441, 18 
\reference{} Malhotra, S., \& Turner, E. L. 1995, \apj\ 445, 553
\reference{} Masetti, N., et al 2000, \aap\ 359, L23 
\reference{} M\'{e}sz\'{a}ros, P., \& Rees, M. J. 1997, \apj\ 476, 232
\reference{} Paczy\'{n}ski, B., \& Rhoads, J. E. 1993, \apjl\ 418, L5
\reference{} Pedersen, H., et al 1998, \apj\ 496, 311
\reference{} Pei, Y. C. 1992, \apj\ 395, 130  
\reference{} Rhoads, J. E. 1997, \apjl\ 487, 1 
\reference{} Sari, R., Piran, T., \& Narayan, R. 1998, \apjl\ 497, 17
\reference{} van Paradijs, J., et al 1997, Nature 386, 686
\reference{} Waxman, E., \& Draine, B. T. 2000, \apj\ 537, 796
\reference{} Williams, R., et al 1996, \aj\ 112, 1335 
\end{references}
\end{document}